\documentclass{appolb}
\usepackage{epsf}
\newcommand{\be} {\begin{equation}}
\newcommand{\ee} {\end{equation}}
\newcommand{\beqa} {\begin{eqnarray}}
\newcommand{\eeqa} {\end{eqnarray}}
\newcommand{\nn} {\nonumber}
\newcommand{\eps}{\epsilon}
\def\P{I\!\!P}
\def\pd{\partial}
\font\fiverm=cmr5
\begin{document}
\pagestyle{plain}
\newcount\eLiNe\eLiNe=\inputlineno\advance\eLiNe by -1
\title{EVOLUTION AT SMALL $x$}

\author{A Donnachie$^a$ and P V Landshoff$^b$ 
\address{$^a$Centre for Mathematical Sciences, Wilberforce Road, \\
Cambridge CB3 0WA\\
pvl@damtp.cam.ac.uk\\ \vskip 5truemm
$^b$Department of Physics, Manchester University\\
Manchester M13 9PL\\
ad@a35.ph.man.ac.uk}}
\maketitle

\begin{abstract}
At present there is no correct theory of evolution 
of $F_2(x,Q^2)$ at small $x$. 
It is a mixture of hard and soft
pomeron exchange and perturbative QCD very
successfully describes the evolution of the
hard-pomeron component. 
This allows the gluon density to be calculated.
It is somewhat different from what is conventionally supposed, but
it leads to a clean PQCD description of the data for
the charm structure function. 
Perturbative QCD breaks down 
for the evolution of the soft-pomeron component of $F_2(x,Q^2)$.
\end{abstract}

\section{Introduction}

The conventional treatment\cite{MRST}\cite{CTEQ} of evolution expands the DGLAP
splitting matrix in powers of  $\alpha_s(Q^2)$. As we will explain, this
is almost certainly wrong at small $x$, and at present we have no correct
theory. However, when we combine PQCD with Regge theory\cite{DL02}, 
this problem
is partially solved and provides a very successful description of data,
not only the complete proton structure function $F_2(x,Q^2)$ but
also\cite{DLGLUON} its charm component $F^c_2(x,Q^2)$. 

\begin{figure}[t]
\begin{center}
\epsfxsize=0.6\hsize\epsfbox[50 50 400 300]{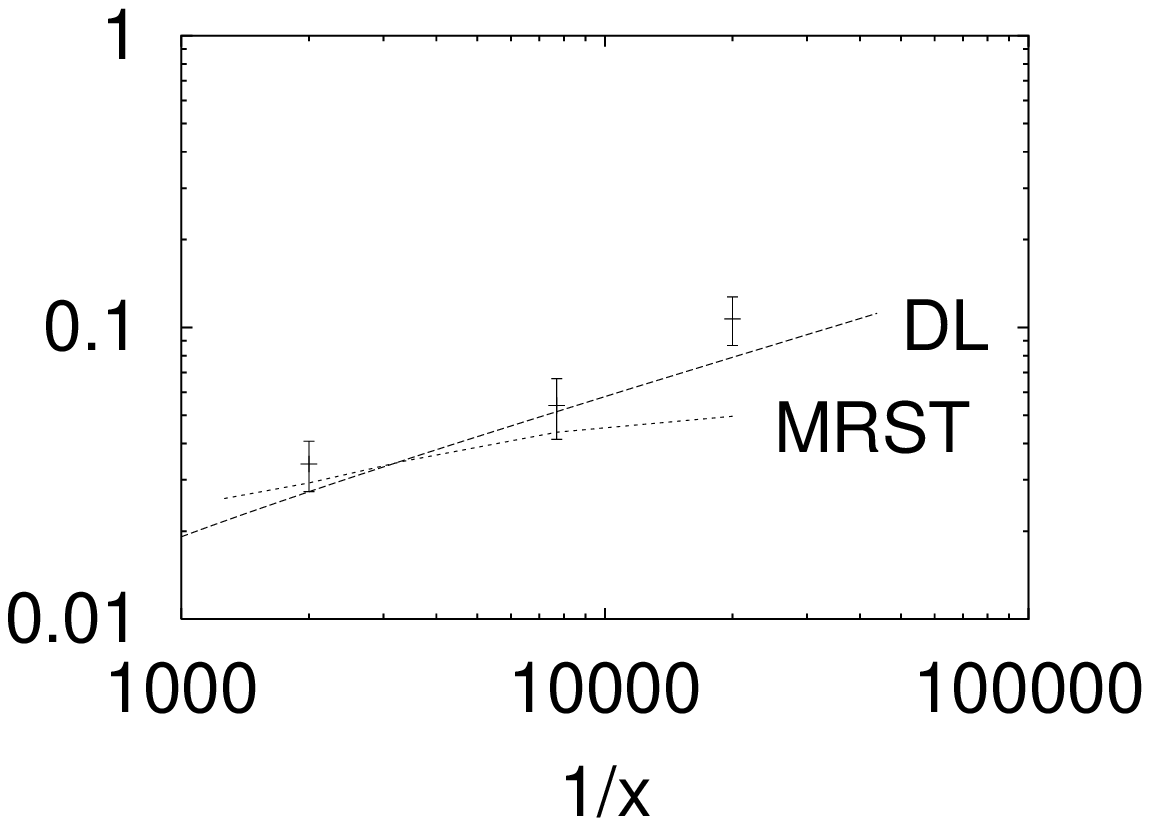}
\end{center}
\caption{Data\cite{ZEUSCHARM} for $F^c_2(x,Q^2)$ at $Q^2=1.8$ GeV$^2$ with 
theoretical curves from references \cite{MRST} and \cite{DLGLUON}}
\label{CHARM}
\vskip 5truemm
\begin{center}
\epsfxsize=0.47\hsize\epsfbox[70 590 300 770]{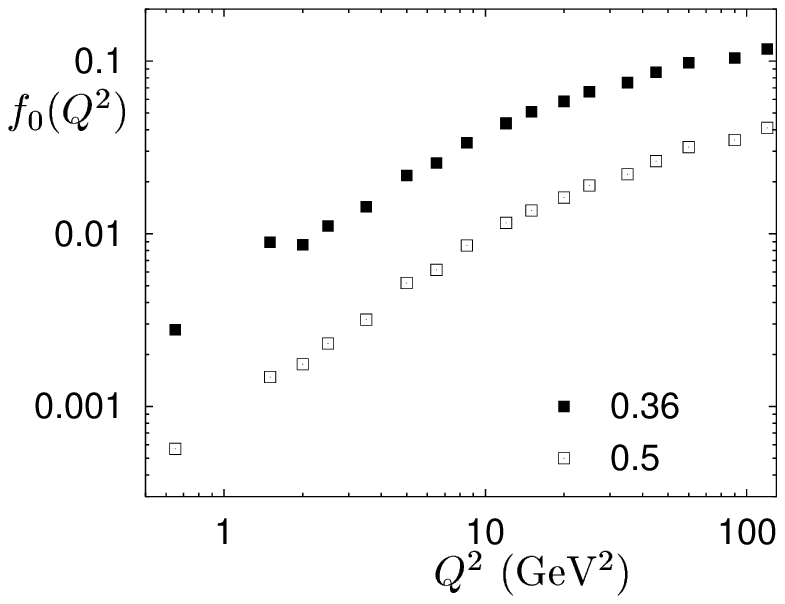}\hfill
\epsfxsize=0.47\hsize\epsfbox[70 590 300 770]{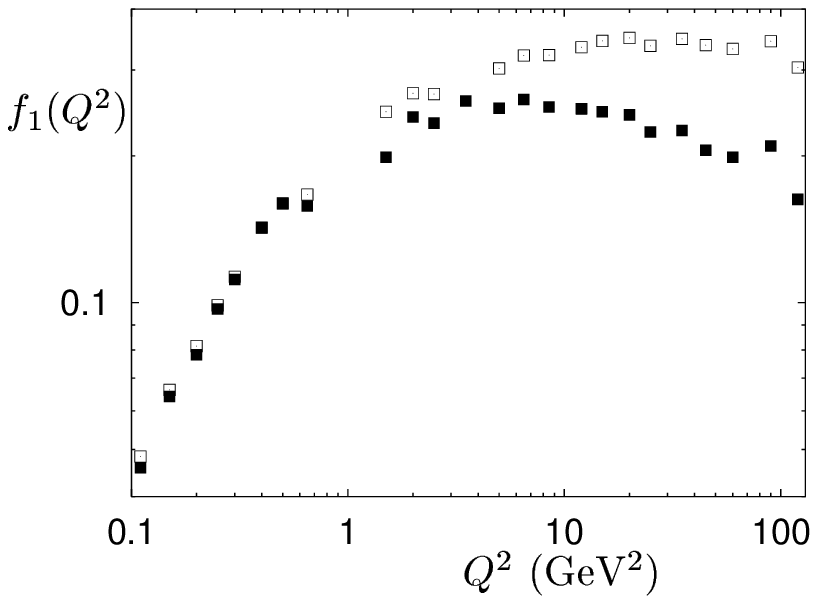}
\end{center}
\caption{The hard and soft pomeron coefficient functions extracted from data}
\label{COEFF}
\end{figure}

The proton's gluon density
is larger at small $x$ than is usually predicted, particularly at small $Q^2$. 
A consequence of this is that PQCD evolution cleanly and
successfully describes 
charm production at small $Q^2$, even down to $Q^2=0$.
See figure \ref{CHARM}. 

\section{Regge theory -- the two pomerons}

At small $x$ we make the fit\cite{DL01}
\be
F_2(x,Q^2)=f_0(Q^2)\,x^{-\eps _0}+f_1(Q^2)\,x^{-\eps _1}
\label{twopom}
\ee
at each $Q^2$ for which there are data. We fix $\eps _1=0.0808$, the
classical soft-pomeron value extracted from hadron-hadron scattering
data\cite{SIGTOT,BOOK}. It turns out that, although the data are now 
highly accurate, they do not constrain the value of $\eps _0$
very closely. Good fits may be obtained with $\eps _0$ anywhere
between 0.35 and 0.5. We call this the ``hard-pomeron'' term.
While varying $\eps _0$ through its allowed range has little effect on the
shape of the hard-pomeron coefficient function $f_0(Q^2)$,
the large-$Q^2$ behaviour of the soft-pomeron coefficient function
$f_1(Q^2)$ changes markedly; see figure \ref{COEFF}.

\begin{figure}
\begin{center}
{\epsfxsize=0.75\hsize\epsfbox[100 300 430 760]{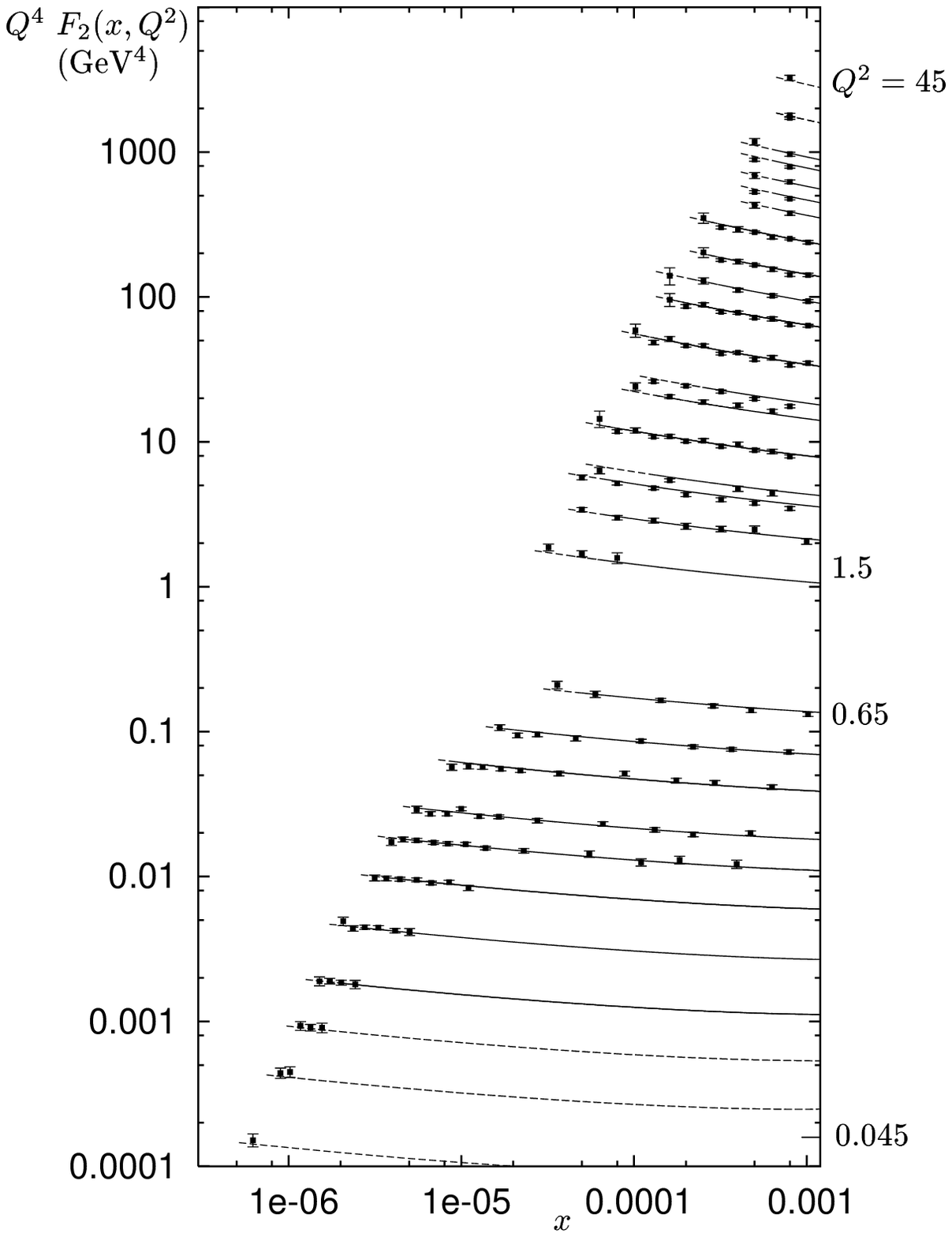}}
\end{center}
\caption{Data from ZEUS\cite{ZEUS}\cite{NEWZEUS} and H1\cite{H1} with two-pomeron fit\cite{DL01}. $Q^2$ ranges from 0.045 to 45 GeV$^2$.}
\label{F2}
\end{figure}

For $\eps _0\approx 0.4$ the data make $f_1(Q^2)$ go to a constant at
large $Q^2$. We assume\cite{DL01} that $f_1(Q^2)$ has this behaviour
and fit the available data for $x\le 0.001$. When we made the fit
we used data from ZEUS\cite{ZEUS} at small $Q^2$ and from H1\cite{H1}
at larger $Q^2$. There are now data at large $Q^2$
also from ZEUS\cite{NEWZEUS}. The best fit is now given by
\be
f_0(Q^2)=A_0\,{(Q^2)^{1+\eps _0}\over(1+Q^2/Q^2_0)^{1+\eps _0/2}}
~~~~~~~~~f_1(Q^2)=A_1\,{(Q^2)^{1+\eps _1}\over(1+Q^2/Q^2_1)^{1+\eps _1}}
\label{coeff}
\ee
with
$$
\eps _0=0.4075~~~\eps _1=0.0808
$$
\be
A_0=0.00227~~~Q_0=2.88\hbox{ GeV}~~~
A_1=0.588~~~Q_1=768\hbox{ MeV}
\label{values}
\ee
See figure \ref{F2}.
We have already explained that the data do not constrain the value of
$\eps _0$ very closely. We have given the parameters to this accuracy
because their errors are strongly correlated. This set of values
gives a $\chi ^2$ per data point significantly less than 1.

\begin{figure}
\begin{center}
{{\epsfysize=\hsize\epsfbox[100 300 410 760]{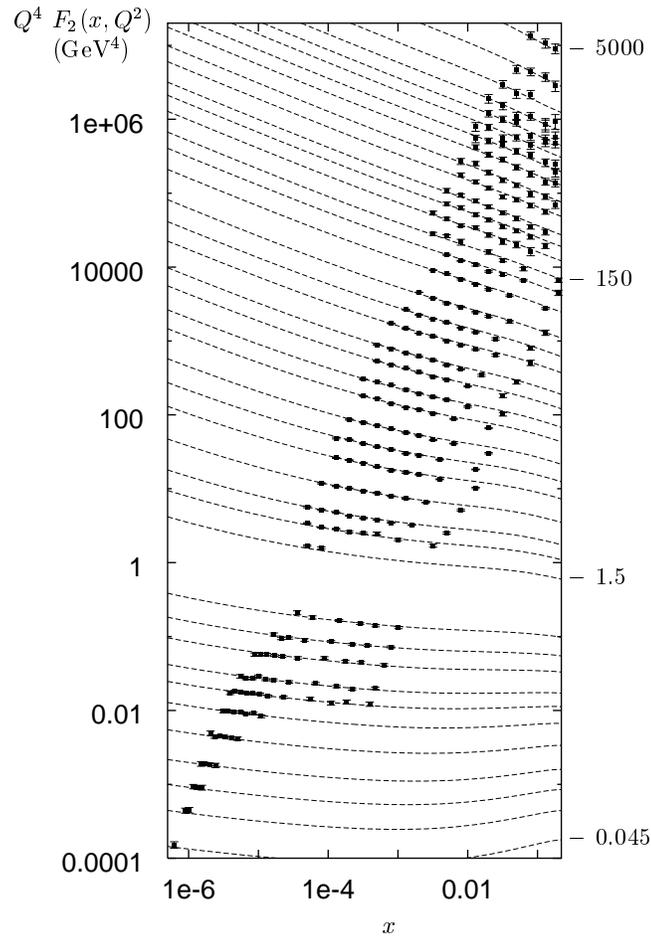}}}
\end{center}
\caption{The fit shown in figure \ref{F2} extended to larger $x$ and $Q^2$}
\label{F2LARGER}
\end{figure}

It is an extremely economical fit: we included in it also data for
photoproduction (figure \ref{PHOTO} below), which largely determine the
value of $A_1$, so that there are just 4 free parameters.
If we multiply (\ref{twopom}) by $(1-x)^7$, which is a {\it very} crude
way of ensuring that $F_2(x,Q^2)$ vanishes as $x\to 1$, and include a term
corresponding to $f_2,a_2$ exchange, the fit agrees well with the data
for larger $x$, even up to $Q^2=5000$ GeV$^2$: figure \ref{F2LARGER}.

\begin{figure}
\epsfxsize=0.3\hsize\epsfbox[90 600 300 765]{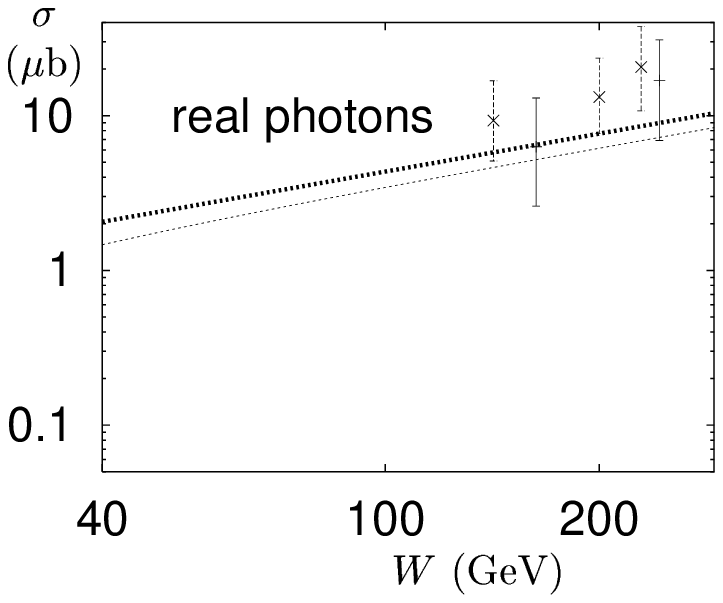}\hfill
\epsfxsize=0.3\hsize\epsfbox[90 600 300 765]{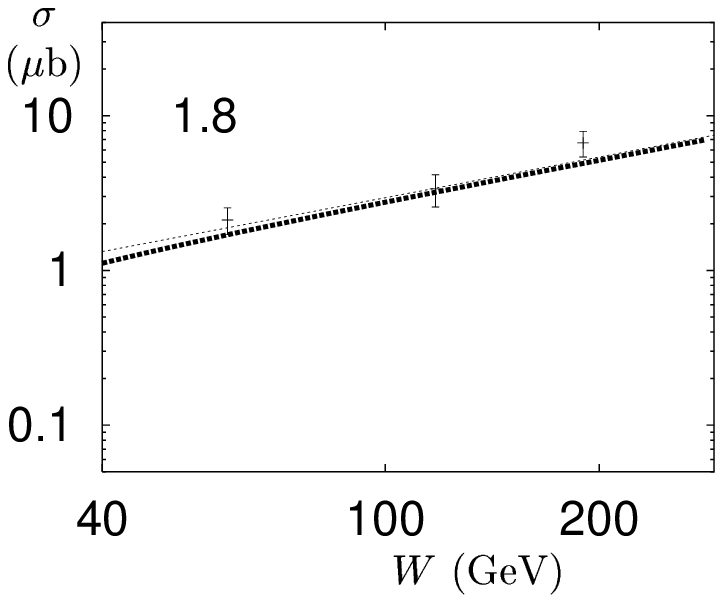}\hfill
\epsfxsize=0.3\hsize\epsfbox[90 600 300 765]{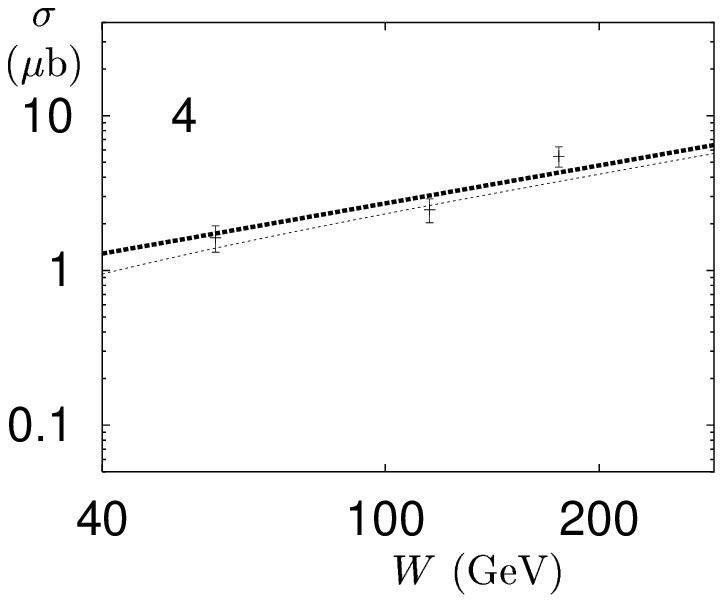}
\vskip 6truemm
\epsfxsize=0.3\hsize\epsfbox[90 600 300 765]{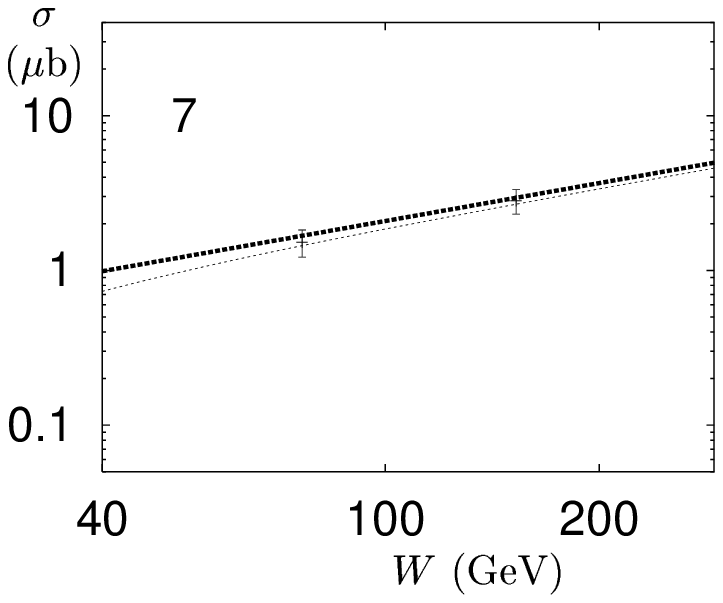}\hfill
\epsfxsize=0.3\hsize\epsfbox[90 600 300 765]{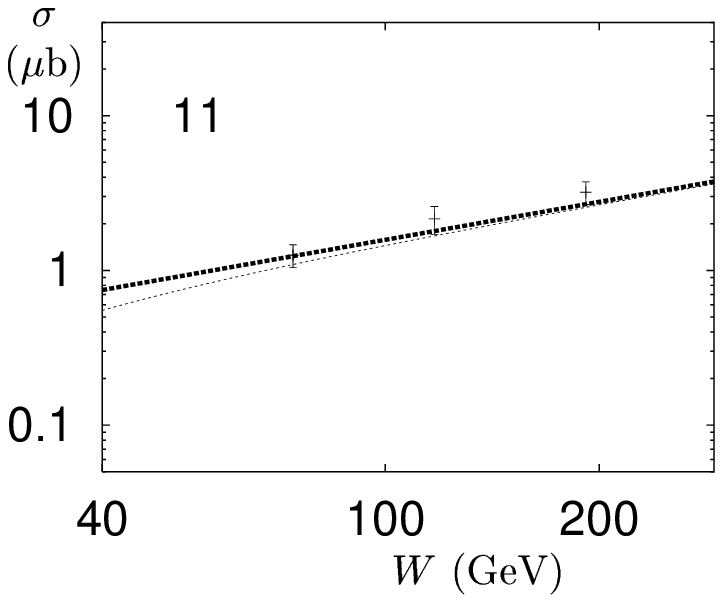}\hfill
\epsfxsize=0.3\hsize\epsfbox[90 600 300 765]{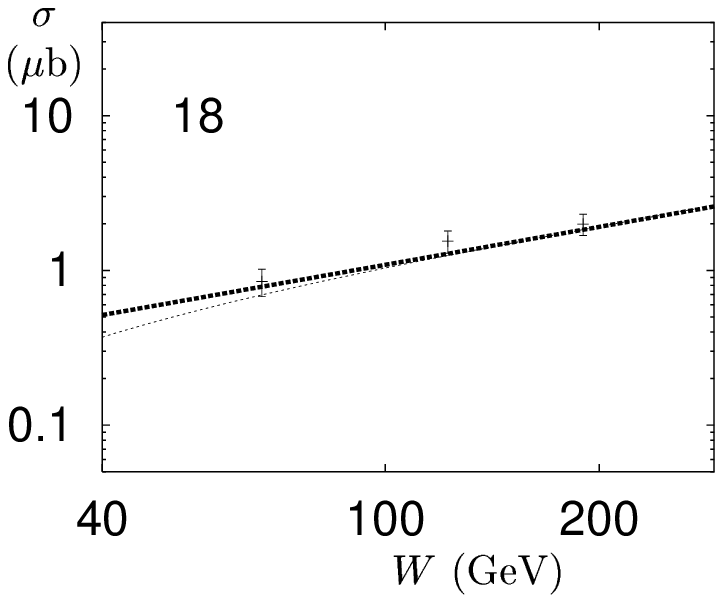}
\vskip 6truemm
\epsfxsize=0.3\hsize\epsfbox[90 600 300 765]{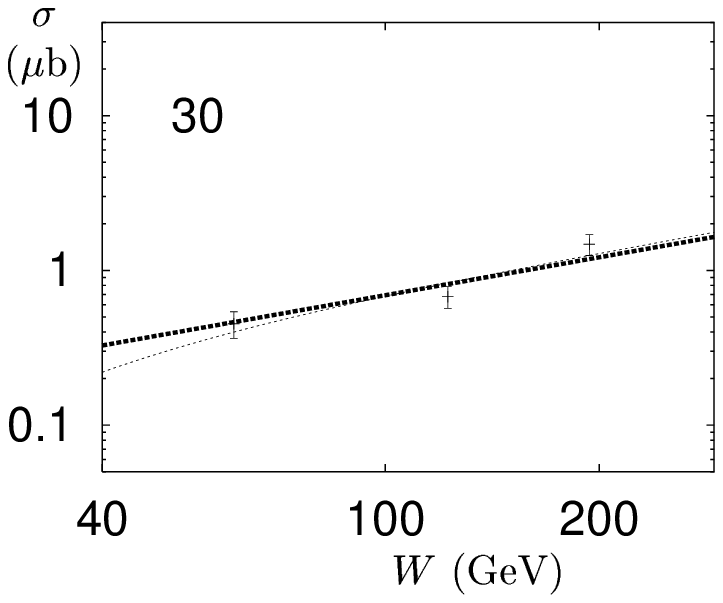}\hfill
\epsfxsize=0.3\hsize\epsfbox[90 600 300 765]{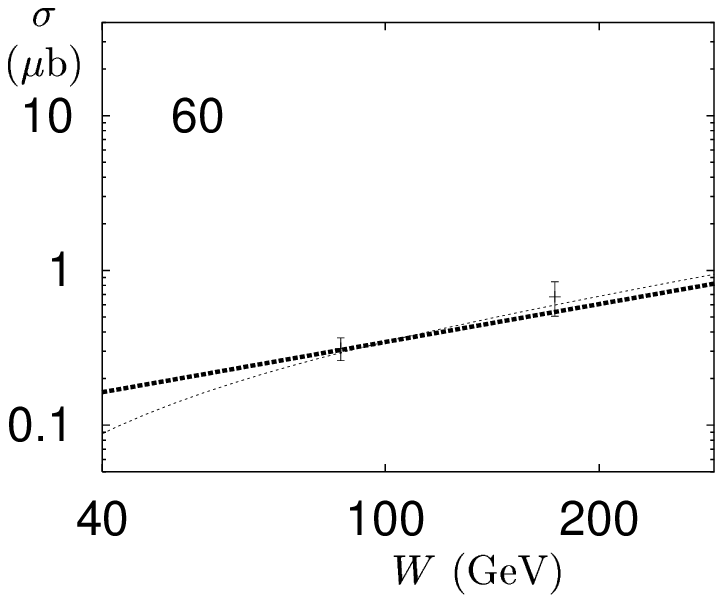}\hfill
\epsfxsize=0.3\hsize\epsfbox[90 600 300 765]{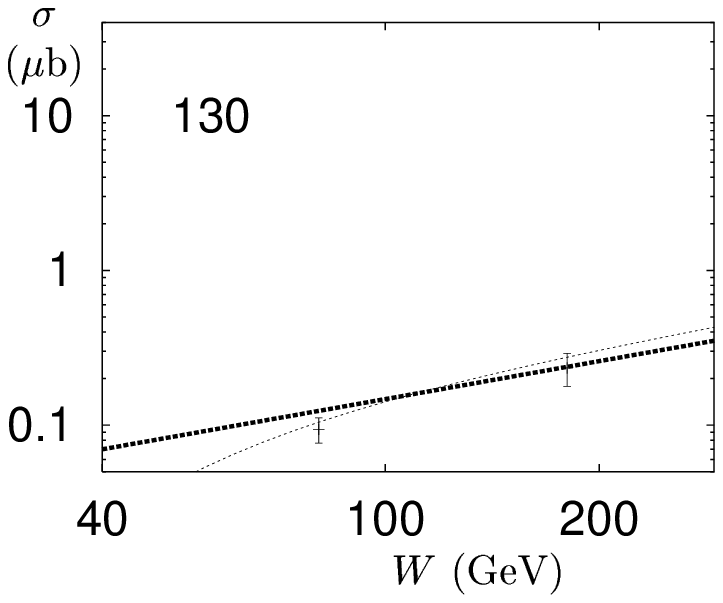}
\caption{Data\cite{ZEUSCHARM} for $\sigma ^c(W)$ defined in (\ref{charmsig}).
The thick lines correspond to (\ref{charmfit}), which coincides with the
output from DGLAP evolution, and the thin lines are
from photon-gluon fusion.}
\label{CHARMSIG}
\end{figure}

If we try making a similar fit to the data\cite{ZEUSCHARM}
for the charm structure function, we find\cite{DL01} that
they correspond only to a hard-pomeron term. Further, the data are
fitted well by assuming that the hard pomeron is flavour blind, so that
for small $x$
\be
F_2^c(x,Q^2)= 0.4\,f_0(Q^2)\,x^{-\eps _0}
\label{charmfit}
\ee
where $f_0(Q^2)$ is defined in (\ref{twopom}). The factor 0.4 is
${4\over 9}/({4\over 9}+{1\over 9}+{4\over 9}+{1\over 9})$.
Figure \ref{CHARMSIG} shows 
\be
\sigma ^c(W)={4\pi^2\alpha_{\hbox{{\fiverm EM}}}\over Q^2}F_2^c(x,Q^2)
\Big|_{x=Q^2/(W^2+Q^2)}
\label{charmsig}
\ee

\section{The DGLAP equation}
\def\u{{\bf u}}
\def\P{{\bf P}}
\def\f{{\bf f}}

Define as usual the singlet parton densities
\be
\u(x,t)= \left (\matrix{x\sum _f(q_f+\bar q_f)\cr xg(x,t)\cr}\right )
~~~~~~t=\log (Q^2/\Lambda^2)
\label{densities}
\ee
and take their Mellin transform with respect to $x$:
\be
\u(N,Q^2)=\int _0^1dx\,x^{N-1}\u(x,Q^2)
\label{mellin}
\ee
Then the DGLAP equation reads
\be
{{\pd\over\pd t}\u (N,Q^2)= \P (N,\alpha_s(Q^2))\,\u(N,Q^2)}
\label{dglap}
\ee
where $\P (N,\alpha_s(Q^2))$ is the Mellin transform of the splitting
matrix.  

The normal procedure is to expand $\P (N,\alpha_s(Q^2))$ in powers of
$\alpha_s(Q^2)$. However, this is \underbar{illegal} when $N$ is close to
0. This is well known. Compare, for example, the analogous expansion of
the function 
\beqa
\psi (N,\alpha_s))&=&\sqrt{N^2+\alpha_s}-N \nn\\
                  &=& \alpha_s/2N-\alpha_s^2/8N^3+\dots
\eeqa
Although each term in the expansion is singular at $N=0$, the function
$\psi$ is not: the expansion is valid only for $|N| > \alpha_s$. 
Similarly, the terms in the expansion of $\P (N,\alpha_s(Q^2))$
have singularities at $N=0$ which are surely not present in
$\P (N,\alpha_s(Q^2))$ itself. Indeed, it is likely that
$\P (N,\alpha_s(Q^2))$ has no relevant $N$-plane singularities at all.

At any given $Q^2$, expanding the splitting matrix in powers of the QCD
coupling becomes invalid when one goes to sufficiently small $x$.

At present, we have no other way to calculate. Luckily, if we introduce
the two-pomeron parametrisation of the data we can partially rescue the
situation. A fixed-power behaviour 
\be
\u(x,t)\sim x^{-\eps}
\label{fixedpower}
\ee
as occurs for each of the terms in (\ref{twopom})
corresponds to
\be
\u(N,Q^2)\sim {\f(Q^2)\over N-\epsilon}~~~~~~~~~~\f (Q^2)=\Big(\matrix{f_q(Q^2)\cr f_g(Q^2)\cr}\Big )
\label{fixedpower2}
\ee
If we insert this behaviour into the DGLAP equation (\ref{dglap})
and equate the coefficient of the pole at $N=\eps$ on each side of the
equation, we find an exact equation that describes how $\f (Q^2)$ evolves
with $Q^2$:
\be
{{\pd\over\pd t}\f (Q^2)= \P (N=\epsilon,\alpha_s(Q^2))\, \f (Q^2)}
\label{evolve}
\ee

\section{DGLAP evolution}

To calculate the evolution of the soft-pomeron term in (\ref{twopom})
we need $\P (N,\alpha_s(Q^2))$ at $N=0.0808$. This is dangerously close
to $N=0$; we cannot make the expansion in powers of $\alpha_s$ and do
not know how to calculate the splitting matrix at
this value of $N$. But for the hard-pomeron
term we need $\P (N,\alpha_s(Q^2))$ for $N\approx 0.4$, which is safely away
from $N=0$ and so the expansion should be valid. 

In order to solve the evolution equation (\ref{evolve}) for the
hard-pomeron coefficient functions $f_q(Q^2)$ and $f_g(Q^2)$,
we chose $Q^2=20$ GeV$^2$ as our starting value. It does
not matter what value we take, as long as it is not too small. We also
assumed that, at this value of $Q^2$, the conventional DGLAP analysis
of the data is correct for values of $x$ down to about 0.01. That is,
we assumed that the value of $g(x=0.01,Q^2=20)$ extracted from the data
by MRST\cite{MRST} or CTEQ\cite{CTEQ} is correct. Further because, as we
have seen, the charm structure function is entirely hard-pomeron exchange
at small $x$ and because, as is well known, it is directly related to
the gluon density, we deduce that $g(x,Q^2)$ at small $x$ is
entirely hard-pomeron exchange. Therefore we know the value of
$f_g(Q^2)$ at $Q^2=20$. For the value of $f_q(Q^2)$ at $Q^2=20$ we
go to our fit to the data, that is we use (\ref{coeff}) and 
(\ref{values}). 

\begin{figure}
\begin{center}
{\epsfxsize=0.47\hsize\epsfbox[75 560 350 765]{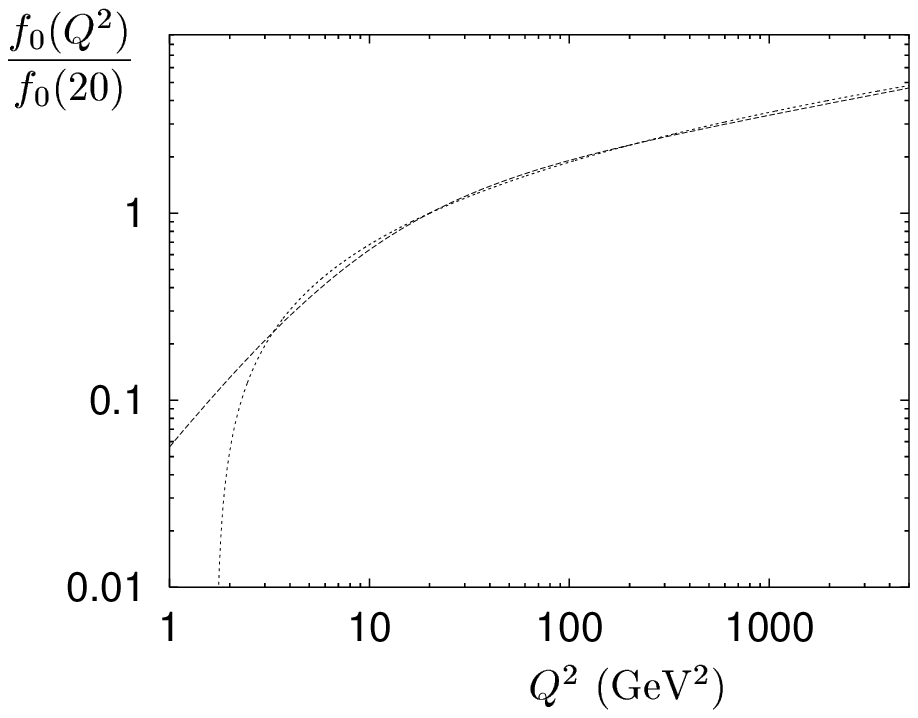}\hfill
\epsfxsize=0.47\hsize\epsfbox[70 560 340 760]{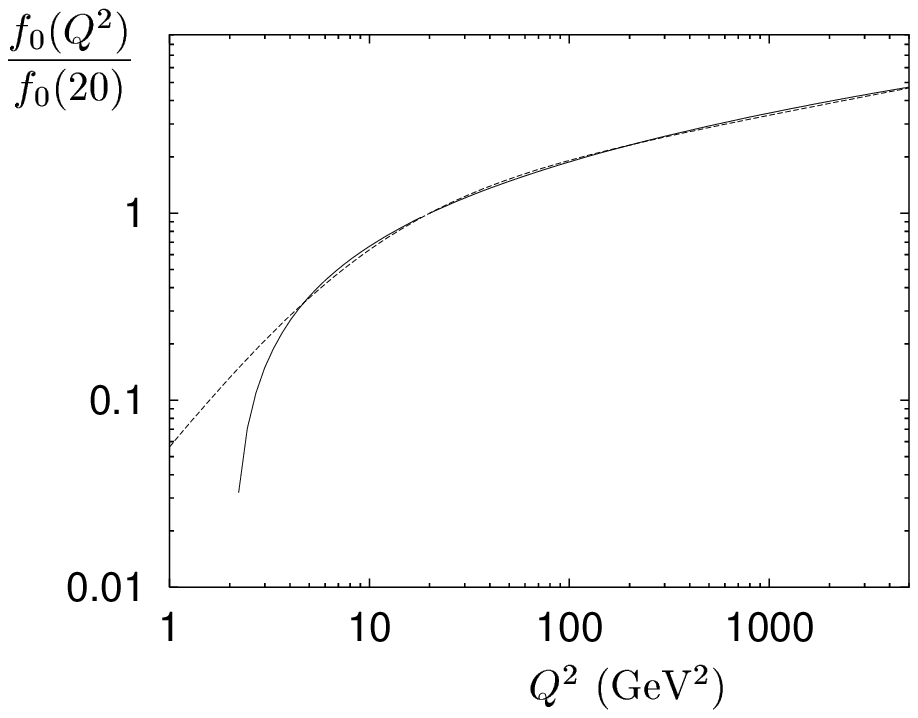}}
\end{center}
\caption{LO and NLO calculations of the hard-pomeron coefficient function,
together with the phenomenological fit (\ref{coeff})}
\label{OUTPUT}
\end{figure}

We then used (\ref{evolve}) to evolve away from $Q^2=20$, in both directions. 
The result is rather astonishing:
although the phenomenological function $f_0(Q^2)$ in (\ref{coeff})
rises at large $Q^2$ as a power of $Q^2$, while the solution to 
(\ref{evolve}) rather rises as a power of $\log Q^2$, the two are
in extraordinarily good numerical agreement over a wide range of $Q^2$.
This is shown in figure \ref{OUTPUT}. 
We took 4 flavours, with $\Lambda_{{\fiverm\hbox{LO}}}=140$ Mev.
We have found\cite{DL02} that the output is the almost same whether we work to
leading order  in the coupling or next-to-leading. 

So, for $Q^2$ greater than about 5 GeV$^2$ perturbative QCD describes
the evolution of the hard-pomeron component of $F_2(x,Q^2)$ extremely
well. This is a significant success both for PQCD and for the two-pomeron
description of $F_2(x,Q^2)$. It is no surprise that perturbative
evolution breaks down at small $Q^2$; the DGLAP equation is supposed to be
valid only for sufficiently large $Q^2$.
 
\section{Gluon density}

\begin{figure}
\begin{center}
{\epsfxsize=0.48\hsize\epsfbox[50 50 390 290]{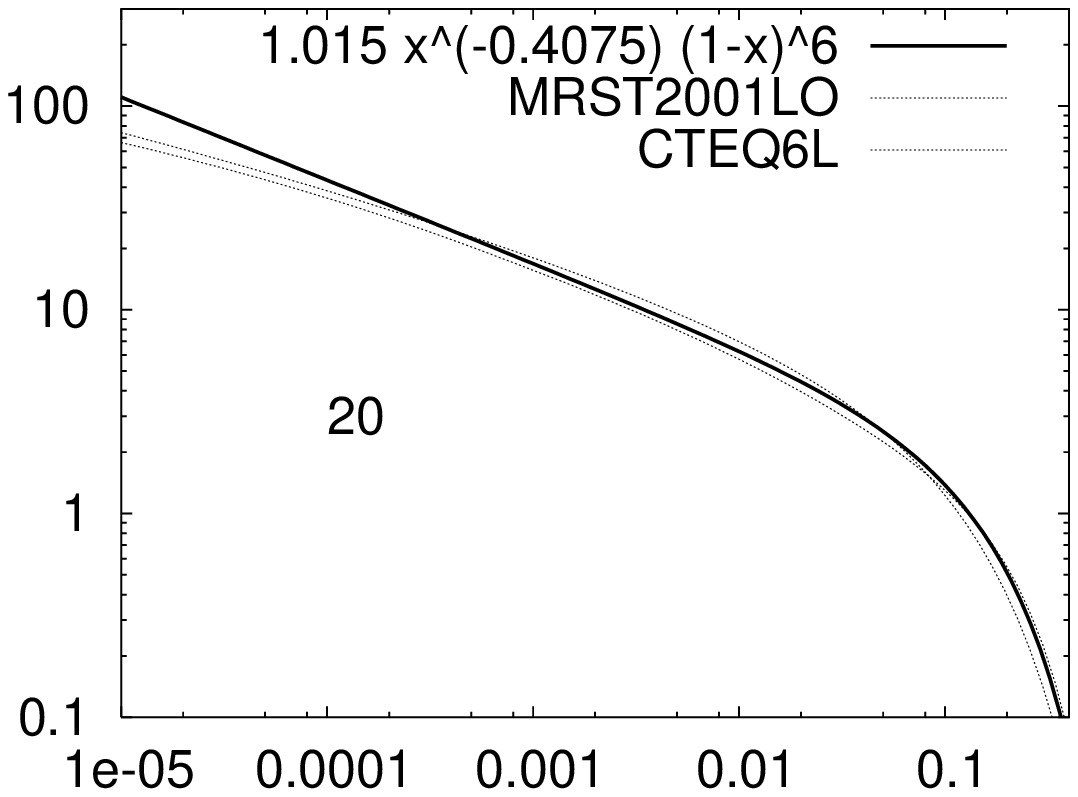}
\hfill
\epsfxsize=0.48\hsize\epsfbox[50 50 390 290]{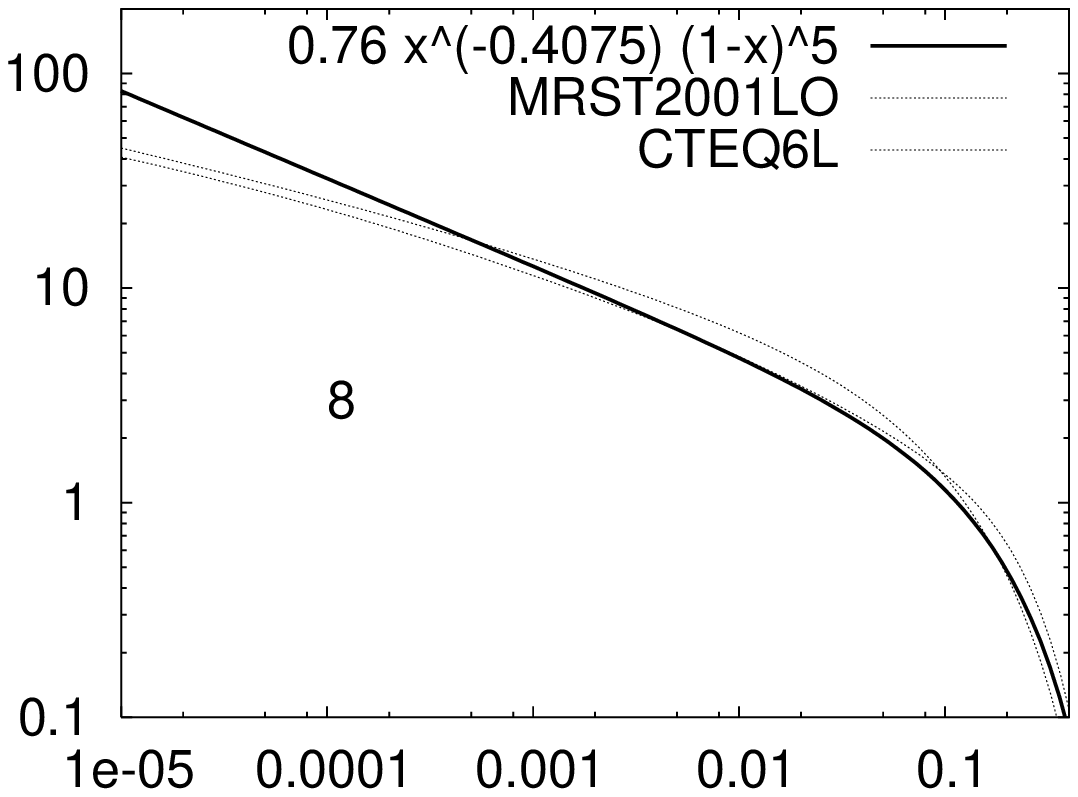}}
\end{center}
\caption{The LO gluon density $xg(x,Q^2)$ at $Q^2=20$ GeV$^2$  and
8 GeV$^2$.
The CTEQ and MRST curves are from the Durham Data Base\cite{DUR}.}
\label{GLUON}
\end{figure}

According to what we have said, the proton's gluon density is
\be
xg(x,Q^2)=f_g(Q^2)x^{-\eps _0}\phi(x,Q^2)
\label{gluon}
\ee
where $\phi(0,Q^2)=1$ and $\phi(x,Q^2)\to 0$ as $x\to 1$. A good numerical
fit to the LO $f_g(Q^2)$ in the range $5<Q^2<1000$ is given by
\be
f_g(Q^2)=0.32 {(Q^2)^{1+\eps _0}\over (1+Q^2/1.4)^{1+\eps _0/2}}
\label{fg}
\ee
At $Q^2=20$ GeV$^2$  
the MRST or CTEQ LO gluon density is
well described by $\phi(x,Q^2=20)=(1-x)^6$.
Figure \ref{GLUON} compares our gluon
distribution in LO with those of MRST and CTEQ at two
values of $Q^2$.  The differences are evident and become even more
pronounced in NLO, where our distribution is much the same but those
of MRST and CTEQ are rather smaller and significantly less steep. 

In NLO, our gluon distribution turns out\cite{DL02} to be almost the same as 
in LO. This is because we use the DGLAP splitting matrix only at
$N\approx 0.4$, where all but one of its elements are almost equal
in LO and NLO. As $x$ decreases, the conventional analysis involves
the splitting matrix at progressively smaller values of $N$, where
its elements are no longer the same in LO and NLO. At very
small $x$  the conventional analysis involves very small $N$, where
it becomes illegal to use the perturbative expansion of the splitting
matrix.

At $Q^2=200$ GeV$^2$ and $x=0.0001$ our gluon distribution is twice
as large as that of MRST or CTEQ.
The fact that our NLO gluon density is larger than the conventional
one at small $x$ will be significant for experiments at the LHC.
 
\begin{figure}
\begin{center}
\epsfxsize=0.48\hsize\epsfbox[50 50 390 290]{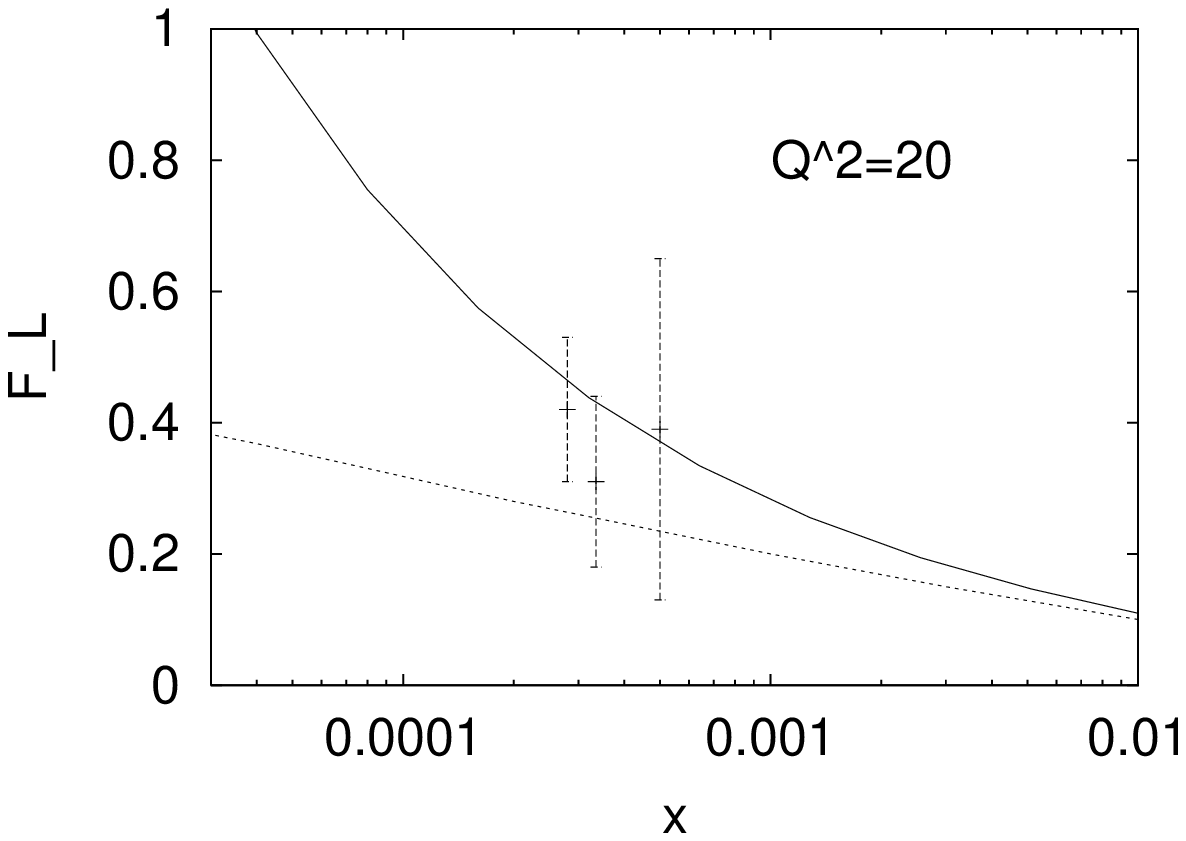}
\end{center}
\caption{The longitudinal structure function at $Q^2=20$ GeV$^2$: data
from H1\cite{h1long} with our prediction (upper curve) and that of
MRST\cite{MRST}}
\label{LONG}
\end{figure}

The cleanest window on the gluon density will be provided by good data
for the longitudinal structure function $F_L(x,Q^2)$. 
Those data that exist depend on some assumed parametrisation
to separate $F_L$ from $F_2$, for example reference \cite{h1long}.
Figure \ref{LONG} shows that already at $Q^2=20$ GeV$^2$ there is
a clear difference between our prediction and that of MRST.

\section{Charm production}

It is standard\cite{acot,robthorne,smith} that at small $Q^2$
the charm structure function $F_2^c(x,Q^2)$ should be calculated
from photon-gluon fusion, $\gamma^*g\to c\bar c$, to some fixed order
in $\alpha_s$. This calculation
introduces some assumed value for the charmed-quark mass $m_c$.
At large $Q^2$ a resummation to all orders in $\alpha_s$ is needed,
because of the presence of  factors of powers of log $Q^2/m^2_c$.
This resummation is achieved by changing at large $Q^2$ to the output
from DGLAP evolution, where the charmed quark mass can now be neglected.
The two calculations have to be matched at some value of $Q^2$. The usual
matching is done at a rather small value of $Q^2$, of the order of
$m_c^2$, and is sensitive to exactly what value is chosen. 

The thin lines in
figure \ref{CHARMSIG} show the result of the LO calculation of 
photon-gluon fusion with our gluon density and $m_c=1.3$ GeV. 
The results of an NLO calculation are almost the same, if
we increase $m_c$ to 1.6 GeV.  This calculation used the code of 
\cite{LSRN,SRN} with our gluon 
distribution and $m_c$ as a free parameter.The thick lines in the figure
are the fit (\ref{charmfit}) which,
as we have shown,  agrees well with the output from DGLAP evolution for
$Q^2$ greater than about 5 GeV$^2$. So in our approach the two
calculations match well over a range of $Q^2$, from about 5 to 50 GeV$^2$.

\section{The hard pomeron}

It used to be a central tenet of high energy physics that scattering
amplitudes are analytic functions of all their variables\cite{elop}.
A consequence of this is that a singlarity that is present in an
amplitude at large $Q^2$ survives when one goes to $Q^2=0$. 
In particular, if the hard pomeron is present at large $Q^2$ it should
also be present at $Q^2=0$. 

This view is strongly reinforced by the charm-production data shown
in figure \ref{CHARMSIG}. The $W$ dependence at $Q^2=0$ is the same
as at higher $Q^2$. The hard pomeron is not \underbar{generated} by PQCD
evolution, though the evolution makes it relatively more important as
$Q^2$ increases. Thus, in the fit we have described to $F_2(x,Q^2)$,
there is a small hard-pomeron component already at small $Q^2$, though
its significance is masked by the much larger soft pomeron component.
Both grow with increasing $Q^2$, but the hard-pomeron component grows
faster, and at each small value of $x$ it dominates at sufficiently
large $Q^2$. 

\begin{figure}
\begin{center}
\epsfxsize=0.7\textwidth\epsfbox[60 580 320 690]{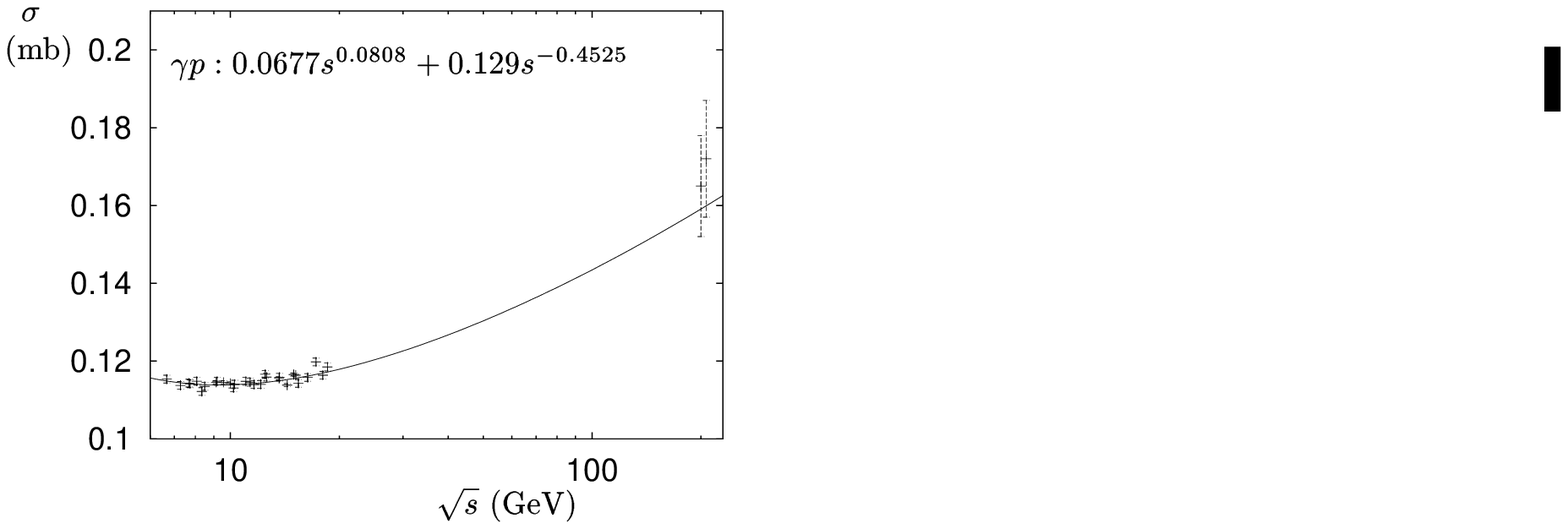}
\end{center}
\caption{$\gamma p$ total cross section; the curve takes account of %
the exchange of the soft pomeron, $f_2$ and $a_2$}
\label{PHOTO}
\end{figure}

Figure \ref{PHOTO} shows the data for the photoproduction cross section.
The curve is our old fit\cite{SIGTOT,BOOK}, with no hard-pomeron term.
The data are not yet good enough to test whether the fit is adequate
or whether an extra component is needed such that indeed the
hard pomeron is present already at $Q^2=0$. The same statement may be
made of the LEP data\cite{OPAL,L3} for $\sigma^{\gamma\gamma}$, 
which depend too heavily on Monte
Carlo simulations that correct for poor acceptance to reach any conclusion.

If the hard pomeron is present in the total cross section for
photon collisons, is the same true for $pp$ collisions? While probably
the hard pomeron couples to a small object such as the photon with
larger relative strength than to a large object such as the proton,
there is some prospect that LHC data will show that there is a hard-pomeron
component to $\sigma^{pp}$.

What is the hard pomeron? Everybody agrees that the sharp rise in
$F_2(x,Q^2)$ at small $x$ discovered at HERA is a consequence of
gluon exchange. Our own belief is that it is caused by glueball
exchange and that the hard and soft pomerons are just glueball
Regge trajectories. There is some evidence that this is true for the
soft pomeron: there is a $2^{++}$ glueball
candidate at 1926 MeV, exactly the right mass to be on the soft-pomeron 
trajectory\cite{WA91}. Another $2^{++}$ glueball candidate\cite{WA102},
at 2350 MeV, could well be on the hard-pomeron trajectory\cite{DL98}.

At one time there was a hope that the power $\eps _0$ of $1/x$, which
is the hard-pomeron-exchange term, might be calculated from the
BFKL and therefore that it is a perturbative effect.
The soft-pomeron trajectory surely cannot be calculated from perturbative
QCD. It may be that the glueballs on the hard-pomeron
trajectory are heavy enough for their masses to be calculated from
PQCD though, with the problems that have arisen with the BFKL
equation,  it is far from clear that PQCD can be used to
calculate the intercept $1+\eps _0$ of the trajectory.

\section{Summary}
{\parindent=0pt

\def\b{{$\bullet ~$}}
\b The conventional approach to evolution needs modifying at small $x$

\b It can be corrected if we combine it with Regge theory

\b But only partly --- we can only treat the hard-pomeron part

\b This is enough to extract the gluon distribution

\b The gluon distribution is larger at small $x$ than has so far been supposed

\b It gives a good description of charm production

\b We want good data for the longitudinal structure function
}

\vskip 10truemm
{\sl This research was supported in part by PPARC}

\end{document}